\documentstyle[11pt,newpasp,twoside,psfig,dcolumn]{article}

\index{pulsars}
\index{pulsars - surveys}
\index{supernova remnants}
\index{pulsars - high-energy emission}

\def\edcomment#1{\iffalse\marginpar{\raggedright\sl#1\/}\else\relax\fi}
\reversemarginpar
\newcolumntype{d}[1]{D{.}{\cdot}{#1}}

\begin{document}
\title{Young Pulsars from the Parkes Multibeam Pulsar Survey and their Associations}
\author{R. N. Manchester, J. F. Bell}
\affil{Australia Telescope National Facility, CSIRO, PO Box 76, Epping
NSW 1710, Australia (Email: Dick.Manchester@csiro.au)}
\author{F. Camilo}
\affil{Columbia Astrophysics Laboratory, Columbia University, 550 West 120th Street, 
New York, NY 10027, USA}
\author{M. Kramer, A. G. Lyne, G. B. Hobbs, B. C. Joshi}
\affil{University of Manchester, Jodrell Bank Observatory, Macclesfield, Cheshire SK11 9DL, UK}
\author{F. Crawford}
\affil{Lockheed Martin Management and Data Systems, PO Box 8048, Philadelphia, PA 19101, 
USA}
\author{N. D'Amico, A. Possenti}
\affil{Osservatorio Astronomico di Bologna, via Ranzani 1, 40127 Bologna, Italy}
\author{V. M. Kaspi}
\affil{McGill University, Ernest Rutherford Physics Building, 3600 University St., Montreal Qc H3A 2T8, Canada}
\author{I.H. Stairs}
\affil{National Radio Astronomy Observatory, Green Bank, WV 24944, USA}

\begin{abstract}
The Parkes multibeam pulsar survey is covering a $10\deg$-wide strip
of the southern Galactic plane from $l=260\deg$ to $l=50\deg$. It
utilizes a 13-beam receiver operating in the 20-cm band on the Parkes
64-m radio telescope and is much more sensitive than any previous
large-scale survey. Most of the 608 pulsars discovered so far are
relatively distant and many are young, with 37 having a characteristic age
of less than $10^5$ years. At least one of these is associated with a
supernova remnant and four other probable associations are suggested.
Several multibeam pulsars have high values of the parameter $\dot
E/d^2$ and are within the position error contours of unidentified
EGRET gamma-ray sources. These possible associations will be tested
with the advent of new gamma-ray telescopes.
\end{abstract}

\section{Introduction}
The Parkes multibeam pulsar survey is a large-scale survey of the
southern Galactic plane from $l=260\deg$ to $l=50\deg$ and with
$|b|<5\deg$ using the Parkes 64-m radio telescope and the multibeam
receiver. This receiver has 13 beams arranged in a hexagonal pattern,
each with dual linear polarization and a bandwidth of 288 MHz centered
on 1374 MHz. The average system temperature is about 21~K,
corresponding to a system-equivalent flux density of about 30
Jy. Signals from each polarization of each beam are filtered to give
96 3-MHz channels, the outputs of which are summed in polarization
pairs, high-pass filtered, integrated and one-bit digitized at
intervals of 250 $\mu$s. The observation time per pointing is 35 min.
Using clusters of workstations at the various collaborating
institutions, data are dedispersed with up to 325 trial dispersion
measures (DMs) and searched for periodic signals. For low-DM pulsars
with periods in the range 10 ms to 5 s, the limiting sensitivity of
the survey is about 0.2 mJy.

The survey commenced in mid-1997 and approximately 95\% of the 2670
pointings required to complete it have been observed.
Most of the data have been processed, resulting in the
discovery of 608 pulsars, including eight binary pulsars and four
millisecond pulsars. The survey and discovery of the first 100 pulsars
are described in some detail by Manchester et al. (2001). Other papers
announcing discoveries of special interest are referenced in that
paper.

\section{Young pulsars from the Multibeam Survey}
\begin{figure}[ht]
\centerline{\psfig{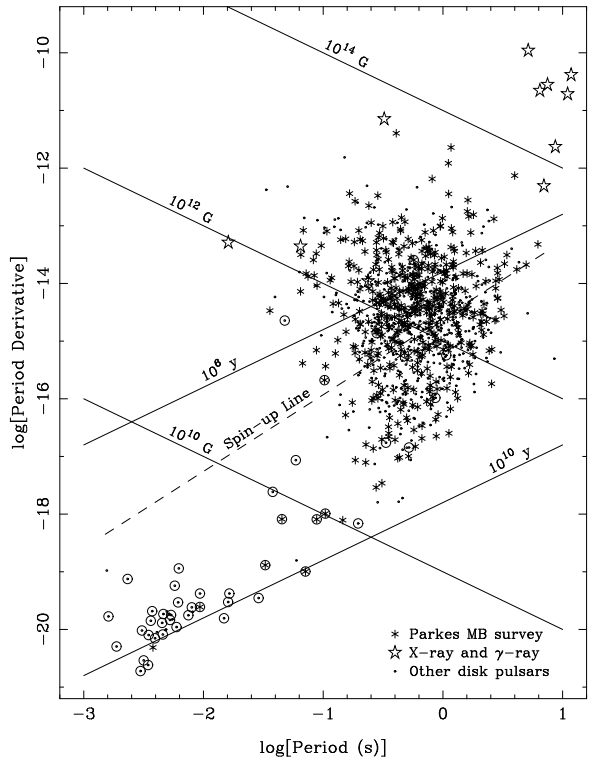}}
\caption{$P-\dot P$ diagram for known disk pulsars, including
anomalous X-ray pulsars (AXPs) and soft $\gamma$-ray repeaters (SGRs)
with known periodicities and slow-down rates. Pulsars discovered in
the Parkes multibeam survey are marked with an asterisk, those 
detected at X-ray or $\gamma$-ray wavelengths only are marked with an open
star and binary pulsars are indicated by a circle around the
symbol. Lines of constant characteristic age $\tau_{\rm c} = P/(2\dot
P)$ and surface dipole magnetic field $B_{\rm s} = 3.2 \times 10^{19}
(P\dot P)^{1/2}$ G, and the spin-up line giving the minimum period
attainable by spin-up due to accretion are marked on
the figure.}
\label{fg:ppdot}
\end{figure}

As shown in Fig.~\ref{fg:ppdot}, the Parkes multibeam survey has been
especially successful in discovering young and highly magnetized
pulsars. In fact, the five radio pulsars with the highest known
surface dipole magnetic fields were all discovered in this survey. Of
the 608 new pulsars, 37 have characteristic ages $\tau_c < 100$ kyr, a
much higher proportion of young pulsars compared to previous
surveys. Surprisingly, about one third of these apparently young
pulsars, including the youngest, PSR J1119$-$6127 (Camilo et al. 2000),
have periods of more than 400 ms, with five exceeding 1 s. A
preliminary `pulsar current' analysis (cf. Lorimer et al. 1993)
assuming a simple disk Galaxy and a beaming factor of five gives a
birthrate of about one pulsar every 180 years, consistent with earlier
estimates.

\section{Supernova Remnant Associations}
With so many newly discovered young pulsars, it is obvious that a
search for associated supernova remnants (SNRs) is likely to be
worthwhile. Following the recognition that PSR J1119$-$6127 is a very
young pulsar, a search of the Molonglo Galactic Plane Survey (MGPS;
Green et al. 1999)\footnote{See
www.astrop.physics.usyd.edu.au/MGPS} showed a faint ring of
emission centered on the pulsar. Radio observations with the Australia
Telescope Compact Array (ATCA) (Crawford et al. 2001) and X-ray
observations with ASCA (Pivovaroff et al. 2001) confirmed this
association.

We have used the ATCA and the MGPS to search for possible SNRs around
the remaining 15 multibeam pulsars with ages less than 50
kyr.\footnote{See www.atnf.csiro.au/research/pulsar/pmsurv/pmpsrs.db}
No data are available for PSRs J0729$-$1448 and J1837$-$0604, both
with ages of about 35 kyr, and there was no obvious association for
PSRs J1702$-$4310 (17 kyr), J1112$-$6103 (33 kyr), J1015$-$5719 (39
kyr), J1637$-$4642 (41 kyr) and J0940$-$5428 (42 kyr). The remaining
eight young pulsars, along with PSR J1119$-$6127, are listed in
Table~\ref{tb:assoc}. This table lists pulsar periods and
characteristic ages, the best estimate of the pulsar distance, in most
cases derived from the DM using the Taylor \& Cordes (1993) Galactic
electron density model, the possibly associated SNR, its apparent
radius, the ratio of the pulsar radial distance from the apparent
center of the SNR to the SNR radius ($\beta$) and the status of the
association.

\begin{table}
\begin{minipage}{120mm}
\caption{Possible Supernova Remnant Associations}
\begin{tabular}{lrd{1}rld{0}d{2}l} \hline
PSR J & \multicolumn{1}{c}{$P$} & \multicolumn{1}{c}{$\tau_{\rm c}$} &
\multicolumn{1}{c}{$d$} & \multicolumn{1}{c}{SNR} &
\multicolumn{1}{c}{$R_{\rm snr}$} & \multicolumn{1}{c}{$\beta$} &
\multicolumn{1}{c}{Status} \\
 & \multicolumn{1}{c}{(ms)} & \multicolumn{1}{c}{(kyr)} &
\multicolumn{1}{c}{(kpc)} & & \multicolumn{1}{c}{(pc)} & & \\ \hline
1119$-$6127 & 407 & 1.6 &$5^*$~ & G292.2$-$0.5 & 12 & 0.00 & Certain \\ 
1357$-$6429 & 166 & 7.3 & 4~~ & G309.8$-$2.6 & 25 & ?    & Possible \\ 
1734$-$3333 & 1169& 8.1 & 7~~ & G354.8$-$0.8 & 21 & 2.2? & Possible \\ 
1420$-$6048 & 68  & 13  & 8~~  & G313.4+0.2   & 33 & 0.2  & Probable \\ 
1413$-$6141 & 285 & 14  &$2^*$~ & G312.4$-$0.4 & 8  & 0.35 & Probable \\ 
1726$-$3530 & 1110& 14  & 10~~ & G352.2$-$0.1 & 8  & 0.0  & Probable \\
1632$-$4818 & 813 & 20  & 8~~  & G336.1$-$0.2 & 35 & 0.15 & Probable \\
1016$-$5857 & 107 & 21  &$3^*$~ & G284.3$-$1.8 & 13 & 1.0  & Possible \\
1524$-$5706 & 1116& 50  & 22~~ & G322.5$-$0.1 & 48 & 0.9  & Unlikely \\
\hline
\end{tabular} 
$^*$ Distance estimate based on SNR
\end{minipage}
\label{tb:assoc}
\end{table}

G309.8$-$2.6 was suggested by Duncan et al. (1997) as a possible
SNR. An ATCA image shows the bright feature southwest of the pulsar
position and weaker emission to the north, but no obvious connection
to the pulsar. 

G354.8$-$0.8 is a shell remnant previously identified by Whiteoak \&
Green (1996). The pulsar lies well outside the remnant and normally
would not be considered a likely association. However, the remnant is
teardrop shaped and pointed directly toward the pulsar with a bright
spot at the point of the teardrop. Furthermore, there is weak evidence
for a larger ring-shaped emission feature on the opposing side. It is
possible that the SNR has a bi-annular morphology (cf. Manchester 1987)
and is larger than previously thought.

PSR J1420$-$6048 lies within and has been associated with a complex
region of radio and X-ray emission sometimes known as the Kookaburra
(D'Amico et al. 2001; Roberts, Romani \& Johnston 2001). The pulsar
lies between the two most prominent emission features, both of which
have a distorted ring shape. It is possible that this is another
example of a bi-annular SNR morphology.

G312.4$-$0.4 is a ring-shaped SNR with two bright regions connected by
a bridge of emission (Whiteoak \& Green 1996). PSR J1413$-$6141 lies
on this bridge of emission, approximately midway between the two
bright regions. This gives the system a morphology very similar to that
of PSR B1509$-$58/G320.4$-$1.2 (Manchester 1987; Gaensler et
al. 1999) and PSR B1338$-$62/G308.8$-$0.1 (Kaspi et al. 1992) and
suggests that collimated winds from the pulsar may be responsible for
the bright regions.

\begin{figure}
\begin{tabular}{cc}
\psfig{file=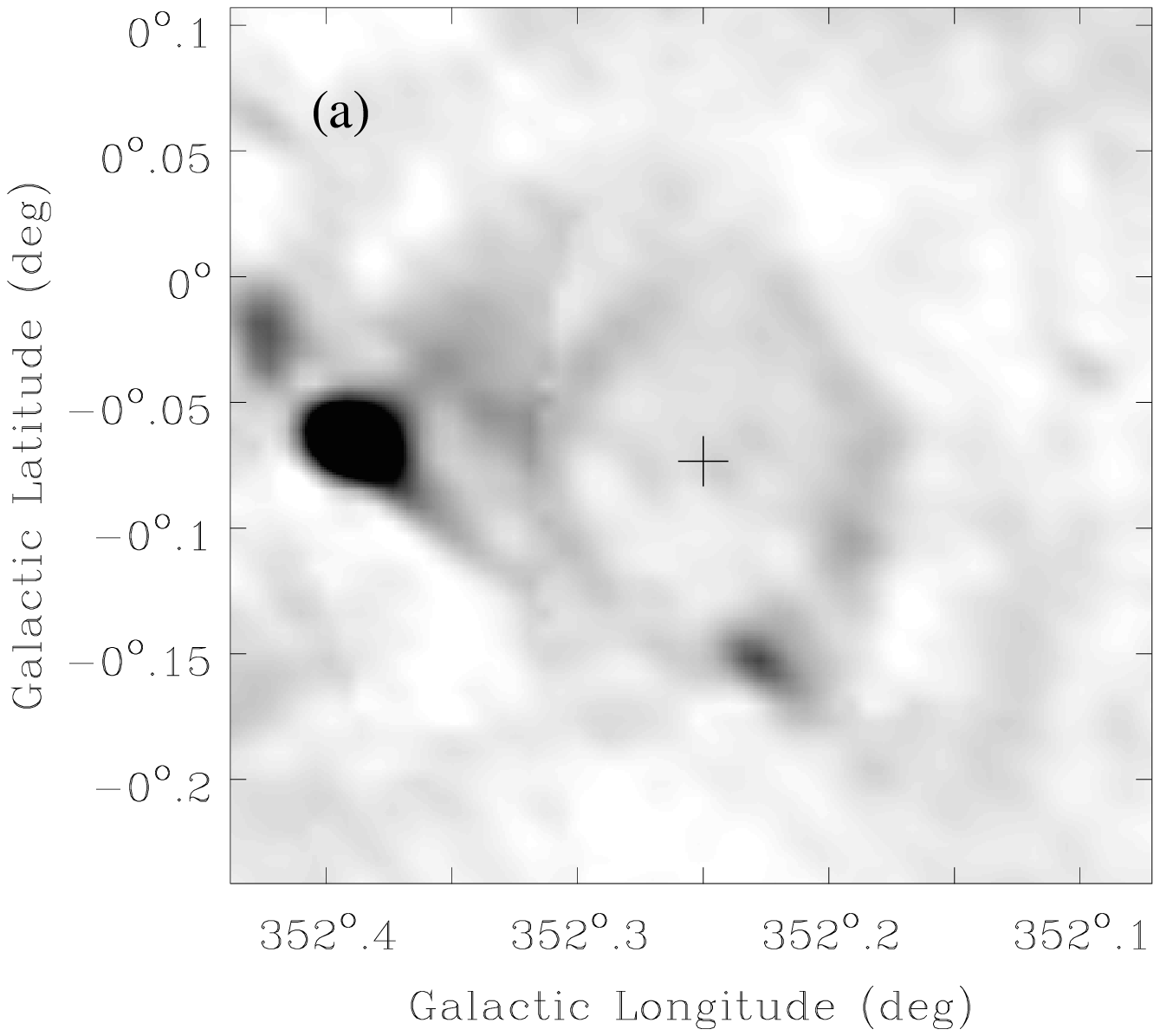,width=62mm} &
\psfig{file=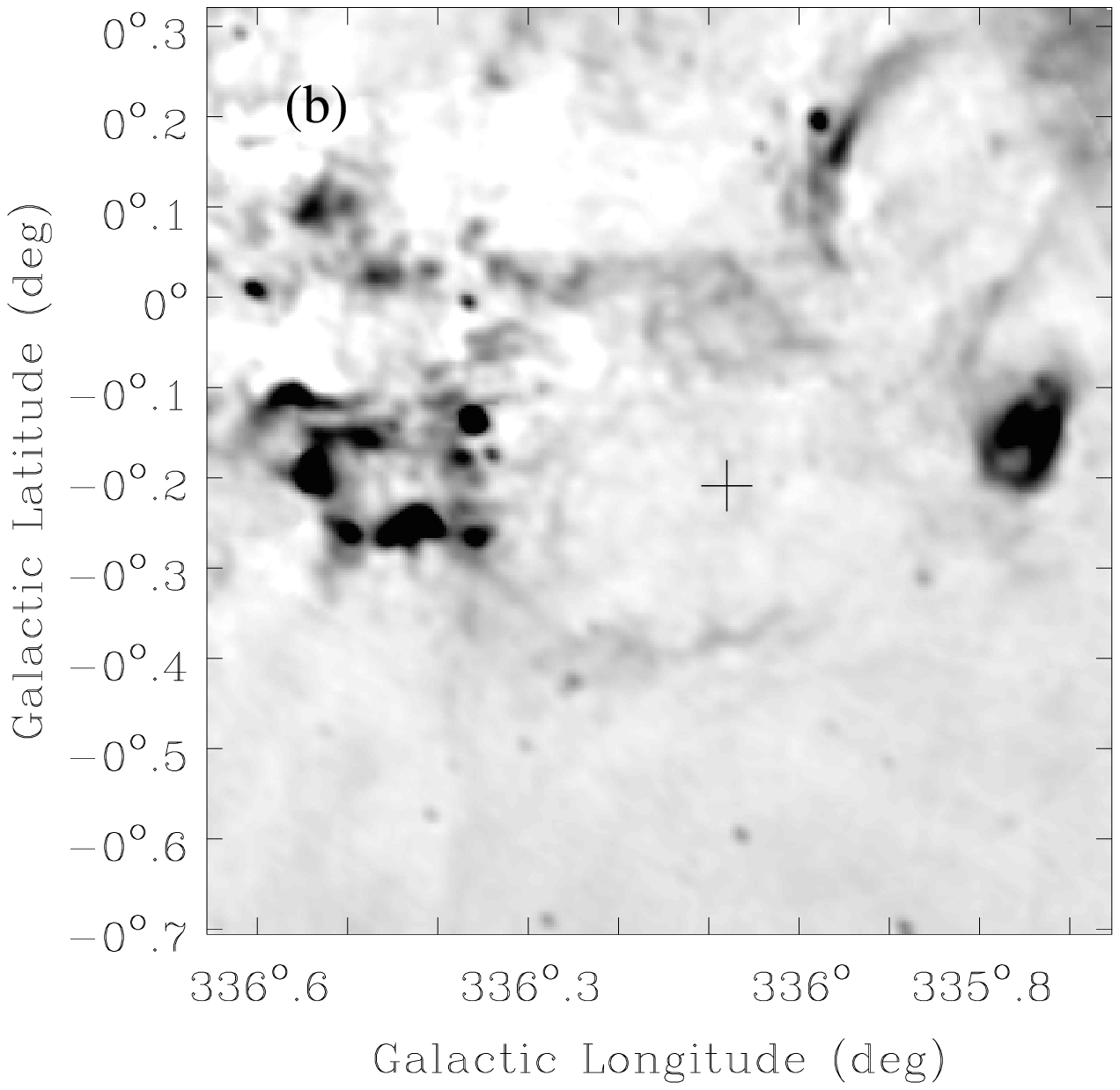,width=60mm} \\
\end{tabular}
\caption{Images from the Molonglo Galactic Plane Survey of probable
supernova remnants surrounding young Parkes multibeam pulsars. The
cross marks the position of the pulsar. (a) G352.2$-$0.1 and PSR
J1726$-$3530.  (b) G336.1$-$0.2 and PSR J1632$-$4818.}
\label{fg:snrs}
\end{figure}

Fig.~\ref{fg:snrs}(a) shows the MGPS image of the region surrounding PSR
J1726$-$3530. The pulsar is located right at the center of a beautiful
shell having bilateral symmetry. This shell, previously uncatalogued,
has been named G352.2$-$0.1. Its morphology and relation to the
pulsar strongly suggest that it is an SNR associated with the
pulsar. Assuming it has the age and distance of the pulsar, its expansion
velocity is a very reasonable 1100 km s$^{-1}$. 

PSR J1632$-$4818 is a 20-kyr-old pulsar which Fig.~\ref{fg:snrs}(b)
shows is located close to the center of another previously
uncatalogued shell source which we have named G336.1$-$0.2. Although
the region is complex, we believe that this shell source is likely to
be an SNR associated with the pulsar. The implied velocity of the
pulsar (250 km s$^{-1}$) and of the shell (1700 km s$^{-1}$) are both
reasonable. 

PSR J1016$-$5857 and its possible association with the SNR
G284.3$-$1.8 have been discussed by Camilo et al. (2001).

PSR J1524$-$5706 lies near the edge of the SNR G322.5$-$0.1 (Whiteoak
\& Green 1996). While the implied pulsar velocity is reasonable, this
SNR has what appears to be a central plerion component, which probably
contains the associated pulsar. Therefore PSR J1524$-$5706 is unlikely
to be associated with this SNR.

In summary, the Parkes multibeam survey has increased the number of
certain or probable pulsar--SNR associations by five, three of these
being with previously unrecognized remnants. It is interesting to note
that SNRs associated with young but long-period pulsars such as PSRs
J1119$-$6127 and J1726$-$3530 are typically difficult to independently
identify. Such objects are likely to remain undetected unless the
associated pulsar is beamed toward us.

\section{Possible Gamma-ray Associations}
Most pulsars known to pulse at $\gamma$-ray energies are young. The
parameter $\dot E/d^2$, where $\dot E$ is the spin-down luminosity and
$d$ is the pulsar distance, is a good indicator of $\gamma$-ray
detectability. Table~\ref{tb:gamma} lists six multibeam pulsars which
are high on a list of all pulsars ranked by $\dot E/d^2$ and which are
located within the error circles (radius $R_{\gamma}$) of unidentified
EGRET $\gamma$-ray sources. The $\beta_{\gamma}$ parameter is defined
analogously to the $\beta$ in Table~\ref{tb:assoc}. The discovery of
the top two on this list, PSRs J1420$-$6048 and J1837$-$0604, and
their possible association with EGRET sources was discussed by D'Amico
et al. (2001). Because most young pulsars sufffer significant period
irregularities, it is not feasible to search the EGRET database for
these pulsars. Confirmation of these associations will have to await
the launch of future $\gamma$-ray telescopes such as AGILE and GLAST.

\begin{table}
\caption{Possible Gamma-ray Associations}
\label{tb:gamma}
\begin{tabular}{lrcrccc} 
\hline
PSR J & \multicolumn{1}{c}{P} &$\tau_{\rm c}$  &  $\dot{E}/d^2$ &  EGRET  & $R_{\gamma}$ & $\beta_{\gamma}$ \\ 
      & \multicolumn{1}{c}{(ms)}& (kyr)  &    Rank    &   3EG   & (deg) & (deg) \\ \hline
1420$-$6048 & 68~ &  13   &   11~   &  J1420$-$6038  &  0.32 &  0.56 \\
1837$-$0604 & 96~ &  34   &   22~   &  J1837$-$0606  & 0.19  &  0.90 \\
1015$-$5719 & 140~ & 39   &   34~   &  J1014$-$5705  &  0.67 &  0.30 \\
1016$-$5857 & 107~ & 21   &   38~   &  J1013$-$5915  & 0.72  &  0.67 \\
1413$-$6141 & 285~ & 13   &   74~   &  J1410$-$6147  & 0.36  &  0.78 \\
1412$-$6145 & 315~ & 50   &  127~   &  J1410$-$6147  & 0.36  &  0.40 \\
\hline
\end{tabular}
\end{table}

\acknowledgments 
RNM thanks Bryan Gaensler for helpful comments and
Malte Marquarding for introducing him to the mysteries of the AIPS++
Viewer. The MOST is operated by the University of Sydney with support
from the Australian Research Council and the Science Foundation for
Physics within the University of Sydney. The Australia Telescope is
funded by the Commonwealth of Australia for operation as a National
Facility managed by CSIRO.

\end{document}